\documentclass[11pt]{article}
\usepackage{amsmath}
\usepackage[russian,english]{babel}

\def\be{\begin{eqnarray}}
\def\ee{\end{eqnarray}}

\makeatletter \let\cl@part\relax \makeatother
\makeatletter \let\cl@chapter\relax \makeatother
\begin{document}
\baselineskip14pt
\bigskip

\begin{center}

\vspace*{0.8cm}

\LARGE{\textbf{ The absence of   QCD  $\beta$-function factorization property
of the generalized  Crewther relation  in  the 't Hooft $\overline{MS}$-based
scheme}}

\vspace*{0.8cm}

\small{\textbf{A.V. Garkusha}}
\footnote{alex.garkusha.92@gmail.com}

\fontsize{7pt}{8pt}

\small{Physics Department, Moscow State University,}

Vorobjevy Gory, 119991, Moscow, Russia

\textbf{A.L. Kataev}
\footnote{kataev@ms2.inr.ac.ru}

Institute for Nuclear Research of the Russian Academy of Sciences,

60th October Anniversary Prospect, 7a, 117312 Moscow, Russia

\vspace*{0.5cm}

\fontsize{8pt}{8pt}

\textbf{ABSTRACT}

\end{center}

We apply the 't Hooft $\overline{MS}$-based scheme to
study  the   scheme-dependence of the QCD   generalization of
Crewther relation
for the  product of  the
normalised non-singlet perturbative contributions
to the $e^+e^-$-annihilation Adler function and to  the  Bjorken sum rule
of  the polarized lepton-nucleon deep-inelastic  scattering process.
We prove  that after  the transformations from
the pure $\overline{MS}$-scheme to the  't Hooft scheme 
the  characteristic   $\overline{MS}$-scheme   theoretical  property
of this
relation, namely the factorization of the  $\beta$-function in its
conformal symmetry breaking part, disappears.  
Another    ``non-comfortable'' theoretical
consequence  of the application
of this prescription  in   $\mathcal{N}=1$ SUSY QED   model is mentioned.
It is shown, that within   
the 't Hooft scheme 
the  expansions  of  Green functions in terms
of the  Lambert   function is simplified in high orders of perturbation theory.
This may be considered as the attractive  
feature  of the 't Hooft scheme, which manifest itself 
in high-order perturbative phenomenological applications.

\newpage

\section{Introduction}

Among fundamental consequences    of the  conformal symmetry in the theory
of strong interactions is the existence of the quark-parton model Crewther
relation
\cite{Crewther:1972kn}, namely
\begin{equation}
\label{Cr}
C_D^{NS} \times C_{Bjp}   = 1~~~.
\end{equation}
Here $C_D^{NS}$ is the normalised Green function, related
to the characteristic of the $e^+e^-$-annihilation to hadrons process, i.e.
non-singlet (NS) part of Adler D-function, defined as
\begin{equation}
D_A^{NS}=\bigg(3\sum Q_f^2\bigg) C_D^{NS}
\end{equation}
where $Q_f$ are the electric  charges of quarks.

The term $C_{Bjp}$ in Eq.(\ref{Cr}) is the 
quark-parton expression for the 
Green function,  which is proportional
to the Bjorken sum rule for the deep-inelastic scattering of polarized leptons on nucleons,
namely
\begin{equation}
S_{Bjp}=\int\limits_0^1 \bigg[g_1^{ep}(x,Q^2)-g_1^{en}(x,Q^2)\bigg]dx=\frac{1}{6}\frac{g_A}{g_V}C_{Bjp}
\end{equation}
Within perturbative QCD both Green functions under consideration can be 
written down  as  
\begin{equation}
\label{CDNS}
 C_D^{NS} (a_s) = 1+ d_0 a_s + d_0 \sum\limits_{i=2}^N d_{i-1} a_s^i 
+O(a_s^{N+1})
\end{equation}
\begin{equation}
\label{CBPS}
C_{Bjp} (a_s) = 1+ c_0 a_s + c_0 \sum\limits_{i=2}^N c_{i-1} a_s^i+O(a_s^{N+1})
\end{equation}
where $a_s=\alpha_s/\pi$ obeys the following renormalization group equation
\begin{equation}
\label{beta}
\mu^2\frac{\partial a_s}{\partial \mu^2}\equiv\beta(a_s)=-\sum_{i\geq 0}\beta_ia_s^{i+2}~~~.
\end{equation}
The application of renormalization procedure, which result in the appearance of the
QCD $\beta$-function,  leads to the violation of the conformal symmetry in QCD
(for a recent review see \cite{Braun:2003rp}) and to the appearance 
in the r.h.s. of Eq.(\ref{Cr}) 
of the additional
conformal symmetry breaking (CSB)  contribution $\Delta_{csb}$: 
\begin{equation}
\label{DNS}
C_D^{NS}(a_s(Q^2)) \times C_{Bjp}(a_s(Q^2))  =  1 + \sum_{i\geq 1}a_s^{i+1}\lambda_i
= 1 + \Delta_{csb}(a_s(Q^2))
\end{equation}
Note that the absence of the $a_s$-term in Eq.(\ref{DNS})
is the consequence of the validity of the variant   of  Eq.(\ref{Cr})
in the
conformal  invariant limit.  In this limit the related equation reads:
\begin{equation}
\label{LQED}
C_{D}^{NS}(a_s)\times C_{Bjp}(a_s)|_{c-i}=1
\end{equation}
This limit is realised e.g. in the case of the
QED approximation, when internal photon vacuum polarization diagrams are 
neglected  \cite{ACGJ}. It is called sometimes as   
perturbative quenched QED (pqQED) model.
The existence of the expression of Eq.(\ref{LQED}) leads to the cancellations
between
the terms proportional to $d_0Q_f^2a$, 
$d_0d_i Q_f^{2(i+1)}a^{i+1}$ in
Eq.(\ref{CDNS})
and the similar terms proportional to $c_0Q_f^2a$, 
$c_0c_i Q_f^{2(i+1)}a^{i+1}$ in Eq.(\ref{CBPS}) (where $a=\alpha/\pi$) in 
the product of these pqQED   Green functions.   
In the  case of QCD these  contributions
to $c_k$ and $d_k$ ($k\geq 0$), which are cancelling due to the 
property of the intial conformal symmetry,   are  labelled by $SU(N_c)$ factors
$C_F^{k+1}$ \cite{Kataev:2008sk}.
One of the consequences of Eq.(\ref{LQED}) is
that the approximations of  Green functions,
which enter into this equation,  are
scale- scheme-independent. This happens in view
of the fact that in the conformal invariant limit  the expansion parameter
does  not depend from   any scale.

Among the consequences of  Eq. (\ref{LQED}) is the following identity:
\begin{equation}
 d_0=-c_0~~~~~.
\end{equation}
It leads to  the absence of the $a_s$-term in Eq.(\ref{CDNS}).

The expression for $\Delta_{csb}$-term in the generalized Crewther
relation is known in the $\overline{MS}$-scheme \cite{Bardeen:1978yd},
which is related to application of the dimensional regularization
\cite{'tHooft:1972fi}. It can be written down as
\begin{equation}
\label{delta}
\Delta_{csb} (a_s(Q^2)) = \Big(\frac{\beta(a_s)}{a_s}\Big)
 K^{\overline{MS}}(a_s);
\end{equation}
where the factorized factor contains the renormalization-group
$\beta$-function, defined in Eq.(\ref{beta}). The polynomial
$K^{\overline{MS}}(a_s)$ has the following form
\begin{equation}
\label{K}
K^{\overline{MS}}(a_s)=a_sK_1^{\overline{MS}}+a_s^2K_2^{\overline{MS}}+
a_s^3K_3^{\overline{MS}}+ \dots
\end{equation}
At the third order of perturbation theory the existence  of this factorized
form of  Eq.(\ref{delta}) was discovered in
Ref. \cite{Broadhurst:1993ru}, where the coefficients
$K_1^{\overline{MS}}$ and $K_2^{\overline{MS}}$ were fixed analytically
using the order $O(a_s^3)$ approximation for $C_D^{NS}$, evaluated
in Ref. \cite{Gorishnii:1990vf} and confirmed in
Refs.\cite{Surguladze:1990tg,Chetyrkin:1996ez},
and the similar perturbative approximation   for $C_{Bjp}(a_s)$, obtained in
Ref.\cite{Larin:1991tj}. Then the validity of the effect of $\beta$-function
factorization
in Eq.(\ref{delta}) was proved in all orders of perturbation theory using the
$\overline{MS}$-scheme \cite{Crewther:1997ux}. The
concrete analytical expression of the
coefficient $K_3^{\overline{MS}}$ was obtained in Ref.\cite{Baikov:2010je}
using the evaluated   $SU(N_c)$ group analytical  expressions
for the   order $a_s^4$ coefficients   in
Eq.(\ref{CDNS}) and Eq.(\ref{CBPS}) in the $\overline{MS}$-scheme
and the
3-loop $\overline{MS}$-scheme expression for
$SU(N_c)$-group $\beta$-function, evaluated
in Ref. \cite{Tarasov:1980au} and confirmed in Ref. \cite{Larin:1993tp}.

It is necessary to remind that the coefficients of the $\beta$-function
do not depend from the concrete realisation of the minimal subtraction
scheme and are the same in the $MS$-scheme  \cite{'tHooft:1973mm},
$\overline{MS}$-scheme, or less known, but rather useful
in practical calculations  $G$-scheme  \cite{Chetyrkin:1980pr}.
However, the coefficients of the perturbation expansions of Green functions
do depend from the concrete realisation of the minimal subtractions
scheme. In our  studies we will use the $\overline{MS}$-scheme  results for
Eq.(\ref{CDNS}),  Eq.(\ref{CBPS}) and the related  expression
for the CSB-term of  Eq.(\ref{delta}).

There is also
another  $\overline{MS}$ representation for the
CSB-term
in Eq.(\ref{DNS}), which is true at the $O(a_s^4)$-level for sure 
and has the following form \cite{Kataev:2010du}
\begin{eqnarray}
\label{delta2}
\Delta_{csb} (a_s)& =& \sum\limits_{n \geq 1}
\Big(\frac{\beta(a_s)}{a_s}\Big)^n
\mathcal{P}_n (a_s) \equiv \sum\limits_{n \geq 1} \sum\limits_{r \geq 1}
\Big(\frac{\beta(a_s)}{a_s}\Big)^n P_n^{(r)} [k,m] C_F^k C_A^m a_s^r
\end{eqnarray}
where  $P_n^{(r)} [k,m]$ with 
 $r+m=k$ do not depend from  Casimir operators. Note, that at this
4-th order of perturbation theory the dependence from
other $SU(N_c)$ group structure constants
$d^{abcd}d^{abcd}$ is absent.

There are   theoretical  questions,
which were  not yet studied  in the case of the representations 
of  Eq.(\ref{delta}) and Eq.(\ref{delta2}) for the  
generalized Crewther 
relation.
One of them is  whether there   is some special theoretical
information, which is encoded
in   Eq.(\ref{delta})  and Eq.(\ref{delta2})
and whether  the factorization  feature is
true in the  $\overline{MS}$-scheme only.   
In this work  for the   analysis of this problem we  apply the 
't Hooft scheme \cite{Hooft}, \cite{'tHooft:1977am}. We  clarify first 
how to use this   scheme more rigorously in
the phenomenologically oriented studies in high orders of
perturbation theory. Next, we will
reveal theoretical problems of the 't Hooft scheme, 
which are manifesting themselves in 
the  studies of the generalized Crewther relation.
Namely, we will demonstrate that the factorization property
of Eq.(\ref{delta}) and Eq.(\ref{delta2}) is not manifesting itself  in the
't Hooft scheme and discuss possible theoretical
explanation of this ``uncomfortable'' result, which may be
compared with the troubles of its application
in pure theoretical   $\mathcal{N}=1$ SUSY QED model.

\section{The key  points of the 't Hooft scheme definition}

It is known that in gauge theories  with single  coupling constant
the first two coefficients of the  $\beta$-function, defined in
Eq.(\ref{beta}), are scheme-independent.
In the   works of Refs. \cite{Hooft}, \cite{'tHooft:1977am} 't Hooft
proposed to use in the theoretical studies the scheme,
which is characterised by $\beta$-function
with two scheme-independent coefficients
only:
\begin{equation}
\label{thooftdef}
\mu^2\frac{\partial a_H}{\partial \mu^2}=\beta(a_H)=-\beta_0 a_H^2 -\beta_1 a_H^3
\end{equation}

Indexes \emph{H} are used for  labelling   't Hooft-scheme parameters. 
The nullification of higher order coefficients
of   the $\beta$-function is
achieved by finite renormalizations of charge.
They are changing the expressions for the
coefficients of  perturbation theory series for Green
functions, evaluated  in the concrete  renormalization scheme,
say $\overline{MS}$-scheme.

This 't Hooft prescription depends on the choice
of the initial scheme,  which is used for the calculation of
a Green function.
In this work we consider massless
perturbative series
for two  Green functions, defined by Eq.(\ref{CDNS}) and Eq.(\ref{CBPS}).
The 
transformed to the 't Hooft
scheme corresponding  perturbative  series can be written down as
\begin{eqnarray}
\label{CDH}
 C_D^{NS}(a_{H})& =& 1+ d_0a_{H} + d_0
\sum\limits_{i=2}^N d_{i-1}^{H}a_{H}^i +O(a_{H}^{N+1})\\
\label{CBH}
C_{Bjp}(a_{H})& =& 1+ c_0 a_{H} +
c_0 \sum\limits_{i=2}^N c_{i-1}^{H}
a_{H}^i+O(a_{H}^{N+1})~~~.
\end{eqnarray}

 Note that,  like in the $\overline{MS}$ -scheme,  
 vector current is  conserved in the 't Hooft scheme.
Indeed, at the 2-loop level the renormalization of the vector 
current is the same as in the $\overline{MS}$-scheme, and therefore its 
divergency  is zero.  In higher orders of perturbation theory 
the conservation of the vector current in the 't Hooft scheme 
holds, since 
higher order corrections to   
to the 't Hooft scheme  $\beta$-function are identically equal to zero.  
Therefore, from this point of view the definition of  this  
scheme is theoretically consistent.

Taking this into account one can safely write 
traditional inverse logarithmic expression   for the
QCD coupling constant $a_{H}$. It reads
\begin{equation}
\label{exp}
a_{H}=\frac{1}{\beta_0 {\rm L}}
-\frac{\beta_1 \ln {\rm L}}{\beta_0^3~{\rm L}^2}
+\frac{\beta_1^2}{\beta_0^5~{\rm L}^3}\big(\ln^2{\rm L}-\ln~{\rm L}-1\big)
-\frac{\beta_1^3}{\beta_0^7 {\rm L}^4}\big(\ln^3{\rm L}+\frac{5}{2}\ln^2{\rm L}
+2 \ln{\rm L}-\frac{1}{2}\big)+ O(\frac{1}{{\rm L}^5})
\end{equation}
where ${\rm L}={\ln(Q^2/\Lambda_{\overline{MS}}^2)}$.
At this stage  instead of the  inverse log expansion of Eq.(\ref{exp})
it is possible to use the expression
through the Lambert functions, which follows
from the explicit solution of Eq.(\ref{thooftdef}) and has
the following form
\begin{equation}
\label{HMS}
a_{H}=-\frac{\beta_0^2}{\beta_1
(1+W_{-1}(z_w({\rm L})))}
\end{equation}
where $z_w(L)=(\beta_0^2/\beta_1){\rm exp}(-1+i\pi-(\beta_0^2/\beta_1){\rm L})$
and $W_k$, $k=0,\pm 1, \dots,$ are the branches of the Lambert function,
defined as the solution of the equation $z=W(z)\exp(W(z))$.
This 2-loop expression  is  applied in the analysis
of multiloop calculations  starting from  the  works
of Refs.
\cite{Magradze:1998ng,Gardi:1998rf,Shirkov:1998sb,Magradze:1999um}
and is used now rather regularly (see e.g. Refs.\cite{Howe:2003mp},
\cite{Bakulev:2006ex}, \cite{Contreras:2010xa}).
Moreover, following the ideas of the works  of Refs.\cite{Howe:2003mp},
\cite{Contreras:2010xa}, we propose  to use Eq.(\ref{HMS}) as the expansion 
parameter 
in the beyond-next-to-leading order QCD studies. However, this step
should be done with care. Indeed, in this case it is necessary
to take into account the
re-calculated to the 't Hooft scheme coefficients   $d_{i}^{H}$ and
$c_{i}^{H}$ in the perturbative series of
Eq.(\ref{CDH}), Eq.(\ref{CBH}) or of any other similar  physical quantities.
In the next section we clarify  how to get them
starting from  the $\overline{MS}$-scheme results and
present their  explicit  expressions up to 4-th order
terms.

\section{Determination of the perturbative expansions  for Green
functions in the 't Hooft scheme}

Consider perturbation series for the Green function of Eq.(\ref{CDH}).
Within the effective-charges approach of Ref. \cite{Grunberg:1982fw}
it can be re-written as
$$C_D^{NS} = 1+ d_0 a_{eff}^D$$
where the effective charge  $a_{eff}^D$ obeys the following
renormalization group equation
\begin{equation}
\mu^2 \frac{\partial}{\partial \mu^2} a_{eff}^{D} = \beta(a_{eff}^{D})=
 -\sum_{n\geq 0}\beta_n^{SI,D}a_{eff}^{D~~(n+2)}
\end{equation}
with scheme-independent (SI),  but Green-function dependent
coefficients $\beta_n^{SI,D}$. They are   related to scheme-invariants,
introduced in Ref.~\cite{Stevenson:1981vj}. Following
the studies of Refs. \cite{Kataev:1994rw}, \cite{Kataev:1995vh},
we  express them in the following form
\begin{equation}
\label{betaSI}
\frac{\beta_n^{SI,D}}{n-1} = \frac{\beta_n}{n-1} +
\beta_0 d_n - \beta_0 \frac{\Omega_n}{d_0}
\end{equation}
where
\begin{eqnarray}
\label{omega2}\Omega_2 &=& d_0 d_1 (\frac{\beta_1}{\beta_0} +d_1) \\
\label{omega3}\Omega_3& =& d_0 d_1 (\frac{\beta_2}{\beta_0}  - \frac{1}{2} d_1 \frac{\beta_1}{\beta_0} - 2 d_1^2 + 3 d_2). \\
\label{omega4}\Omega_4& =& \frac{d_0}{3} (3 d_1 \frac{\beta_3}{\beta_0}+ d_2 \frac{\beta_2}{\beta_0}- 4 d_1^2 \frac{\beta_2}{\beta_0}+
2 d_1 d_2 \frac{\beta_1}{\beta_0}- d_3 \frac{\beta_1}{\beta_0}+ 14 d_1^4 - 28 d_1^2 d_2 + 5 d_2^2 + 12 d_1 d_3).
\end{eqnarray}
It is possible to find the expressions
for ${\Omega_n}$ (with $n\geq 4$) as well ( ${\Omega_n}$ was obtained
in Ref.\cite{Kataev:1994rw}). At the present level
of the development of calculations machinery it is enough to stop
at this level. In fact taking into account $\Omega_4$-expression
is already related to the not  yet achieved in QCD   5-th perturbative 
level. However,
its consideration  will help to reveal some interesting features.

To get the expressions for the coefficients $d_n$ in the 't Hooft
scheme we  use Eq.(\ref{betaSI}) and write down the following
identity:
\begin{equation}
\label{main}
\frac{\beta_n^{\overline{MS}}}{n-1} + \beta_0 d_n^{\overline{MS}} -
\beta_0 \frac{\Omega_n^{\overline{MS}}}{d_0} =
\frac{\beta_n^{H}}{n-1} + \beta_0 d_n^{H} -
\beta_0 \frac{\Omega_n^{H}}{d_0}
\end{equation}
Taking into account that within 't Hooft scheme
\begin{equation}
\beta_n^{H} \equiv 0~ {\rm at}~~  n \geq 2
\end{equation}
and using Eq.(\ref{omega2})-Eq.(\ref{omega4}),  we find the following
expressions for the coefficients of Green functions:
\begin{eqnarray}
\label{recalc1}
d_0 d_2^{H}& =&
d_0 d_2^{\overline{MS}} + d_0 \frac{\beta_2^{\overline{MS}}}{\beta_0} \\
\label{recalc2}
 d_0 d_3^{H}& =& d_0 d_3^{\overline{MS}} + \frac{d_0}{2} \frac{\beta_3^{\overline{MS}}}
{\beta_0} + 2 d_0 d_1^{\overline{MS}} \frac{\beta_2^{\overline{MS}}}{\beta_0} \\
\label{recalc3}
d_0 d_4^{H}& =& d_0 d_4^{\overline{MS}} +
\frac{1}{3} d_0 \frac{\beta_4^{\overline{MS}}}{\beta_0} +
d_0 d_1^{\overline{MS}} \frac{\beta_3^{\overline{MS}}}{\beta_0} +
3 d_0 d_2^{\overline{MS}} \frac{\beta_2^{\overline{MS}}}{\beta_0} +
\frac{5}{3} d_0 \big(\frac{\beta_2^{\overline{MS}}}{\beta_0} \big)^2 -
\frac{1}{6} d_0 \frac{\beta_1 \beta_3^{\overline{MS}}}{\beta_0^2}~~.
\end{eqnarray}
Absolutely identical formulae can be obtained for the  coefficients $c_i$ of 
Eq.(\ref{CBPS}) for  the  Bjorken polarized sum rule 
 coefficient function $C_{Bjp}$. It  is possible to use in the series of 
Eq.(\ref{CDH}) and Eq.(\ref{CBH})
the exact coupling constant  expression through Lambert function
from Eq.(\ref{HMS}) after taking into account the evaluated  correction terms. 
Therefore, on the
first glance, the application of the 't Hooft scheme has attractive features
of the simplification of the  perturbative studies within several approaches,
and in particular within  the one proposed in Ref. \cite{Shirkov:1997wi}
(for a detailed review see Ref.\cite{Bakulev:2008td}).

\section{The generalized Crewther relation and theoretical problems of
the 't Hooft scheme}
To find  
the structure of the generalized Crewther relation in the
$\overline{MS}$-version of the 't Hooft scheme it is
necessary to construct the product
$C_D^{NS}(a_{H})C_{Bjp}(a_{H})$
of Eq.(\ref{CDH}) and Eq.(\ref{CBH})
with taking into account the explicit expressions of the obtained above  
coefficients
$d_0d_2^{H}$,  $d_0d_3^{H}$ (see Eq.(\ref{recalc1}) and
Eq.(\ref{recalc2}))
and the similar expressions for the coefficients $c_0c_2^{H}$, 
$c_0c_3^{H}$.
Using them together with the  results for
$d_0d_1^{\overline{MS}}$
\cite{Chetyrkin:1979bj,Dine:1979qh,Celmaster:1979xr},
$d_0d_2^{\overline{MS}}$
\cite{Gorishnii:1990vf,Surguladze:1990tg,Chetyrkin:1996ez},
$d_0d_3^{\overline{MS}}$ \cite{Baikov:2010je} and for the similar
$\overline{MS}$-scheme
 terms in  $C_{Bjp}$, obtained in
Refs.\cite{Gorishnii:1985xm,Zijlstra:1992kj}, Ref.\cite{Larin:1991tj}
and Ref.\cite{Baikov:2010je}, we come to  the following statement:

{\bf Statement.} In  the  't Hooft $\overline{MS}$-based scheme
there is no explicit factorization of the  terms
$(\beta(a_H)/ a_H)$ in the QCD  generalizations of
Crewther relations of Eq.(\ref{delta}) and Eq.(\ref{delta2}). This
feature
distinguishes it from $\overline{MS}$-scheme (or from   any other version of
$MS$-scheme) and is raising the questions of  applicability of
the 't Hooft scheme for revealing  theoretical effects, hidden
in analytical high-order corrections to Green functions and $\beta$-function
as well.

{\bf Proof.} Let us compare the following representations
for the CSB term in the generalized Crewther
relation of Eq.(\ref{DNS})\footnote{For the sake of generality,
we included also  non-calculated  $a_s^5$ $\overline{MS}$-scheme
corrections.}:
\begin{eqnarray}
\label{MSbar}
\Delta^{\overline{MS}}_{csb}(a_{\overline{MS}})& =&
(-\beta_0 a_{\overline{MS}} - \beta_1 a^2_{\overline{MS}} -
\beta_2^{\overline{MS}} a^3_{\overline{MS}}
-\beta_3^{\overline{MS}}a^4_{\overline{MS}} )\times \\
\nonumber
&& \times (a_{\overline{MS}} K_1^{\overline{MS}} +
a_{\overline{MS}}^2 K_2^{\overline{MS}} +
a^3_{\overline{MS}} K_3^{\overline{MS}}+
a^4_{\overline{MS}} K_4^{\overline{MS}}) \\ \label{Hooft}
\Delta^{H}_{csb}(a_H)&=&
(-\beta_0 a_H - \beta_1 a_H^2)(a_H K_1^H + a_H^2 K_2^H + a_H^3 K_3^H
+a_H^4 K_4^H)~~~.
\end{eqnarray}
The analytical expressions for
 $K_1^{\overline{MS}}$,
$K_2^{\overline{MS}}$ in Eq.(\ref{MSbar}) were obtained  in
Ref. \cite{Broadhurst:1993ru}, while the expression for
$K_3^{\overline{MS}}$ was found in Ref.\cite{Baikov:2010je}.
The possibility to add unknown term  $K_4^{\overline{MS}}$ to
the second part of Eq.(\ref{MSbar}) follows from the general
proof of Ref. \cite{Crewther:1997ux} of the validity of 
perturbative all-orders   factorization
of $(\beta(a_s)/a_s)$ factor  in the  CSB
part of the $\overline{MS}$-scheme generalization of the   Crewther relation.

 Multiplying   the series for  $C_D^{NS}(a_{H})$
and $C_{Bjp}(a_{H})$ fom Eq.(\ref{CDH}) and Eq.(\ref{CBH})
and  taking into account transformation formulae of
Eq.(\ref{recalc1}), Eq.(\ref{recalc2}), and Eq.(\ref{recalc3})
for both $d_0d_i^H$ and $c_0c_i^H$-terms,  we find  the
expressions  for $K_1^{H}$, $K_2^{H}$, $K_3^{H}$, $K_4^{H}$-terms.
The results read
\begin{eqnarray}
 K_1^H &=& K_1^{\overline{MS}} \equiv K_1 \\
 K_2^H &=& K_2^{\overline{MS}} \equiv K_2 \\
K_3^H &=&
K_3^{\overline{MS}} + 3 K_1 \frac{\beta_2^{\overline{MS}}}{\beta_0} \\
K_4^H &=&
K_4^{\overline{MS}} + 4 K_2 \frac{\beta_2^{\overline{MS}}}{\beta_0}
+ 2 K_1 \frac{\beta_3^{\overline{MS}}}{\beta_0}
\end{eqnarray}
where the coefficients  $\beta_2^{\overline{MS}}$ and
$\beta_3^{\overline{MS}}$ of the QCD
$\beta$-function  are known from calculations of
Refs.\cite{Tarasov:1980au,Larin:1993tp}
and Refs.\cite{vanRitbergen:1997va,Czakon:2004bu} correspondingly.
Note that in QCD  two latter equations cannot be simplified.
There is no polynomials in Casimir operators in the terms 
$K_3^{H}$, $K_4^{H}$, etc.  
Further on, 
we can rewrite the expression for Eq.(\ref{Hooft}) in the following
form
\begin{eqnarray}
 \Delta_{csb}^{H}(a_H)& =& \Big(-\beta_0 a_H - \beta_1 a_H^2\Big)
\Big( K_1 a_H + K_2 a_H^2 + K_3^{\overline{MS}} a_H^3 + K_4^{\overline{MS}}
a_H^4 \Big) - \\ \nonumber
&&
- 3 \beta_2^{\overline{MS}} K_1 a_H^4  -
a_H^5 \Big[\beta_0 K_4^{\overline{MS}} + 4 K_2 \beta_2^{\overline{MS}} +
2 K_1 \beta_3^{\overline{MS}} +
3 K_1 \frac{\beta_1 \beta_2^{\overline{MS}}}{\beta_0} \Big] + O(a_H^6)
\end{eqnarray}

One can see that there is
no explicit factorization in 't Hooft renormalization procedure.
It can be shown that the similar conclusion is true
for another representation of the generalized Crewther relation of 
Eq.(\ref{delta2}) as well.

\section{Discussions}
One may  wonder what might be the reason for the drastic
differences between the forms of the generalized Crewther
relation in the $\overline{MS}$-scheme and the  't Hooft scheme.
To our point of view the failure  to reproduce factorized form
of the $\overline{MS}$-scheme generalization of the  Crewther
relation is connected with the absence of  diagrammatic representations
of the corresponding results in the~ 't Hooft scheme. Indeed,
while within $\overline{MS}$-scheme it is possible to understand
the origin of the factorized form of Eq.(\ref{MSbar}) (see e.g.
Ref.\cite{Crewther:1997ux}), it seems impossible to formulate
on the diagrammatic level the origin of the appearance
of non-factorizable corrections to the generalized Crewther relation
in the 't Hooft prescription. The similar problem arises if one
tries to use it,  say, in
$\mathcal{N}=1$ SUSY QED model.
Indeed, it was shown in Ref.\cite{Stepanyantz:2011jy}
that the application in  this  model of the  
based on the covariant derivative regularization approach
\cite{Slavnov:1971aw}, \cite{Slavnov:1972sq}
SUSY invariant regularization \cite{Krivoshchekov:1978xg},
allows to understand on the diagrammatic language
the origin of the existence of the scheme with
$\beta$-function, expressed through anomalous dimension 
of superfield \cite{Novikov:1983uc}. In the 't Hooft prescription
of Eq.(\ref{thooftdef}) this feature 
will be never seen. This observation, together with the
fail to reproduce factorizable expression for the $\overline{MS}$-scheme
variant of Crewther relation leads to the conclusion that
one should not use 't Hooft prescription in the theoretical studies
of the special features of gauge theories, which are manifesting themselves
in the renormalization-group calculations, performed at the beyond-two-loop
level. In principle, this was foreseen  by 't Hooft himself
in the work of Ref.\cite{Hooft}, where he wrote ``We do think
perturbation theory up to two loops is essential to obtain accurate
definition of the theory. But we were not able to obtain sufficient
information on the theory to formulate self-consistent procedure
for accurate computations''. This statement makes quite
understandable the doubts on applicability of the 't Hooft
scheme in asymptotic regime \cite{Suslov:2006iy}, related to
applications of the perturbative results  obtained
beyond-two-loop approximation.

This forgotten statement of Ref.\cite{Hooft}  is also
supporting our non-comfortable feeling from  
the fail to reproduce factorizable expressions of the Crewther
relation, explicitly obtained  with the help of application
of  another 't Hooft prescription - the scheme of minimal
subtractions for subtracting ultraviolet divergences
\cite{'tHooft:1973mm} at the level of order $a_s^4$ corrections
\cite{Baikov:2010je}, \cite{Kataev:2010du}.

{\bf Acknowledgements.}
One of us (AVG) is  grateful to G. 't Hooft for
his elucidations of historical aspects and arguments which
led him to introducing the procedure called now by his name.
ALK is grateful to D.V. Shirkov and  D.I. Kazakov for encouraging
to study the problem of scheme-dependence of Crewther
relation  and to  M.J.C. Veltman
for his help to get at hands the original work of Ref.\cite{Hooft}
two decades ago.
Both of us are grateful to K.V.Stepanyantz for interest to the work
and useful comments. The work of one of us (AVG) comes from his
Bachelor Thesis, done in Physics Department of Moscow State
University. The work of ALK was supported in part  by the RFBR grants
No 11-01-00182, No 11-02-00112  and the grant NS-5525.2010.2.

 \end{document}